\documentclass[final,1p,times]{elsarticle} 

\usepackage{graphicx}
\usepackage{amssymb} 
\usepackage{amsthm} 
\usepackage{lineno}
\usepackage{wrapfig}

\newcommand \ee {$e^+e^-$ }

\newcommand \pt {p$_T$ }

\newcommand \sqn {$\sqrt{s_{_{NN}}}$ }
\newcommand \sq {$\sqrt{s}$ } 
\newcommand \dau {d+Au }
\newcommand \auau {Au+Au }
\newcommand \sqnr {$\sqrt{s_{_{NN}}}$~=~200~GeV}

\newcommand \raa {R$_{AA}$ }

\def\NIM{Nucl. Instr. and Meth.}
\def\NIMA{{Nucl. Instr. and Meth.}~{\bf A}}

\def\PLB{{Phys. Lett.}~{\bf B}}

\def\PRL{Phys. Rev. Lett.\ }

\def\PRD{{Phys. Rev.}~{\bf D}}
\def\PRC{{Phys. Rev.}~{\bf C}}

\begin{document}

\begin{frontmatter} 

\title{Photons and low-mass dileptons: results from PHENIX}

\author{Itzhak Tserruya  (for the PHENIX\fnref{col1} Collaboration)}
\fntext[col1] {A list of members of the PHENIX Collaboration and acknowledgements can be found at the end of this issue.}
\address{Weizmann Institute of Science, Rehovot, Israel}


\begin{abstract} 
Most recent PHENIX results on electromagnetic probes are presented including first preliminary results obtained with the Hadron Blind Detector (HBD) on \ee invariant mass spectra from \auau collisions at \sqnr.     
\end{abstract} 

\end{frontmatter} 


\section{Introduction}
The potential 
of electromagnetic probes (real and virtual photons) to characterize the strongly interacting quark gluon plasma
is well known since it was first pointed out by Shuryak \cite{shuryak}. They are unique in providing information about the plasma temperature, the in-medium properties of hadrons and the mechanism of chiral symmetry restoration \cite{gale-rapp-tserruya}. The PHENIX detector was designed with special emphasis on the measurement of such probes \cite{phenix-detector} and has produced the first RHIC results on dileptons \cite{ppg088}  and direct photons \cite{ppg086-photons-auau}. Capitalizing on the flexibility of the RHIC facility, PHENIX is engaged in a systematic study of electromagnetic probes. In these proceedings, I present the most recent PHENIX results on these probes, including the nuclear modification factor \raa and the elliptic flow v$_2$ of direct photons in \auau collisions, the measurement of direct photons in \dau collisions and the first results obtained with the HBD on the measurement of \ee pairs in \auau collisions at \sqnr.

\section{Nuclear modification factor \raa of direct photons}
\begin{figure*}[t]
     \begin{center}
           \includegraphics*[width=66mm, height=43mm] {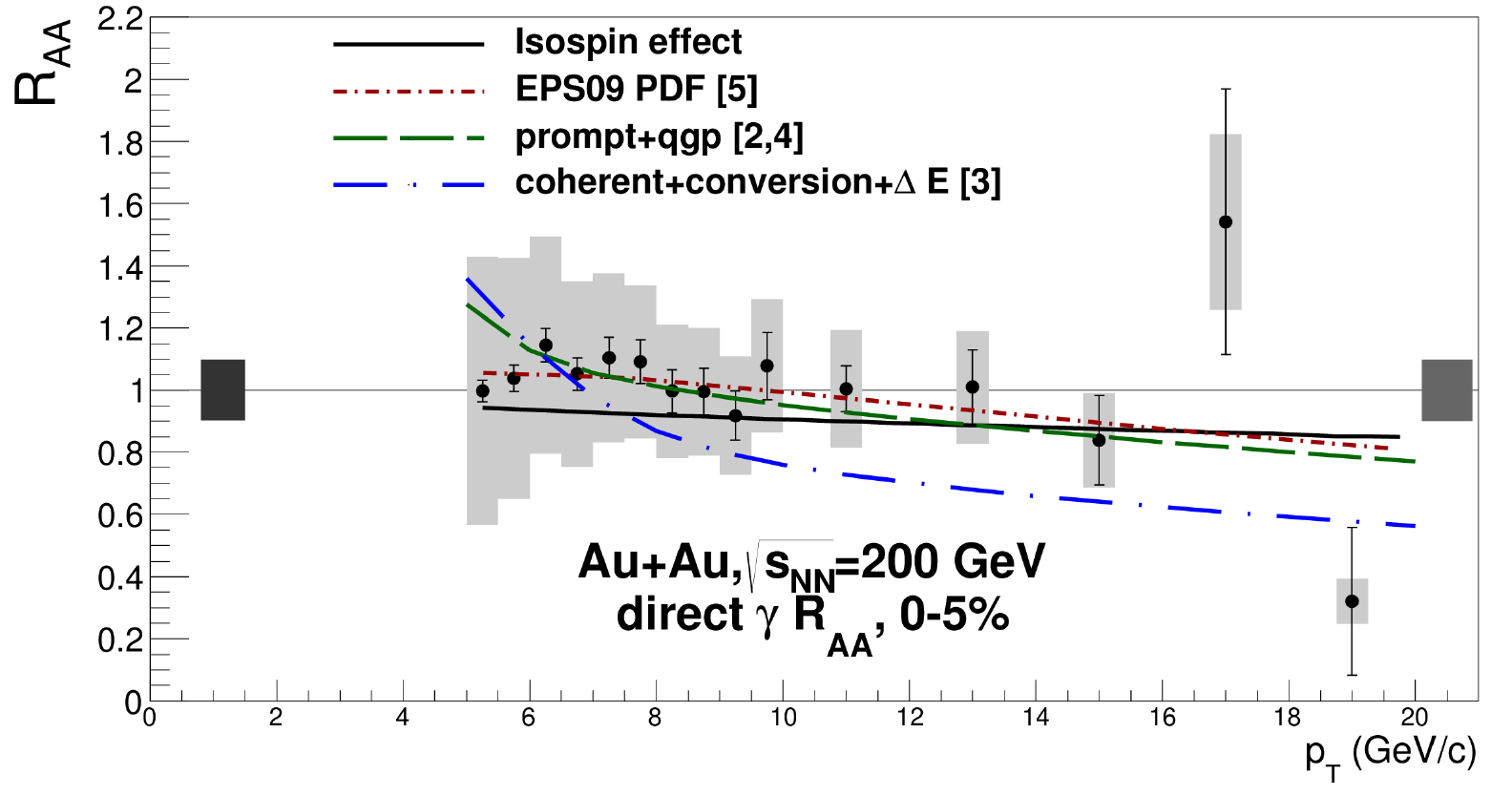}
           \includegraphics*[width=66mm] {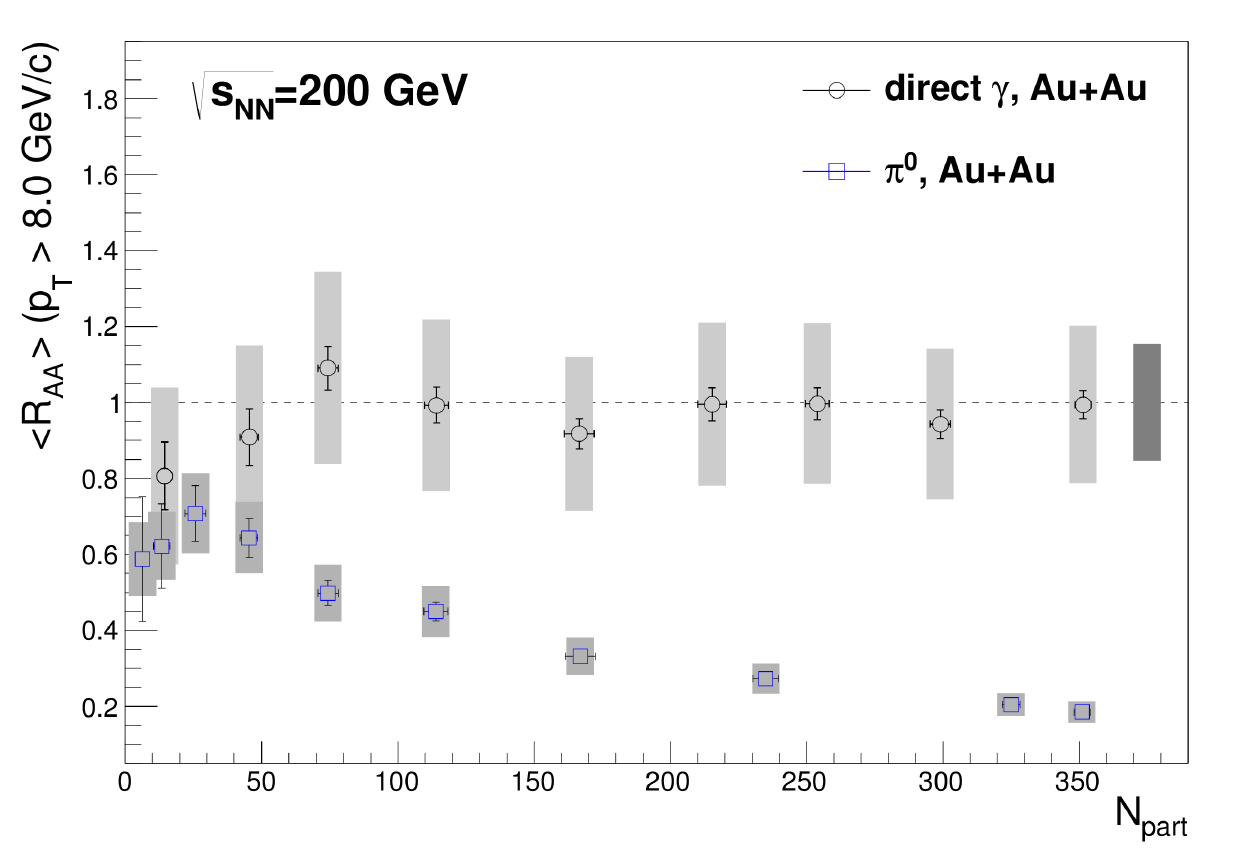}
\vspace{-3mm}
\caption{Left: \raa vs. \pt of direct photons in central \auau collisions at \sqnr.  Right: \raa vs. N$_{part}$ of direct photons and $\pi^0$ for p$_T>$ 8 GeV/c in \auau collisions at \sqnr. For more details about the data and curves see \cite{ppg139-raa-photons}. }
           \label{fig:raa-photons}
     \end{center}
\vspace{-5mm}
\end{figure*}

The \raa of direct photons up to \pt = 20 GeV/c was first reported by the PHENIX experiment \cite{prel-raa-photons}. The preliminary results showed an intriguing suppression trend at high p$_T$, above $\sim$ 12 GeV/c, albeit with very large statistical and systematic uncertainties, that triggered discussions about possible initial state and isospin effects \cite{initial-state-isospin-effects}. To qualitatively improve this result, the \raa was recalculated using the statistically improved 2006 p+p data \cite{ppg136-pp2006} together with a  reanalysis of the \auau data that combines the information of both the PbGl and PbSc calorimeters. The final results presented in  Fig. \ref{fig:raa-photons} show that \raa is consistent with 1 up to \pt = 20 GeV/c and for all centralities expressed by the number of participant nucleons N$_{part}$ \cite{ppg139-raa-photons} thus resolving the long-standing issue of the behavior of the direct photons \raa at high \pt. 

\section{Direct photons in \dau}
PHENIX has measured direct photons in p+p and \auau collisions at \sqnr  \cite{ppg086-photons-auau}. The p+p yield is well reproduced by NLO pQCD calculations down to \pt=1 GeV/c, whereas the \auau data exhibit a strong excess of photons (in the \pt range of 1-3 GeV/c) beyond the  p+p yield scaled by the number of binary collisions N$_{coll}$. This excess has an exponential shape with a slope parameter  T = 221 $\pm 19^{stat} \pm 19^{syst}$ MeV in central (0-20\%) collisions and has been interpreted as thermal radiation from the medium thus providing the first information about the temperature of the system averaged over the space-time evolution of the collision. Using hydrodynamical models one can infer an initial temperature of T$_{ini}$ = 300 to 600 MeV depending on the assumed formation time ($\tau$ = 0.6-0.15 fm/c) of the system. 

\begin{figure*}[h!]
    \begin{center}
           \includegraphics*[width=65mm, height=70mm] {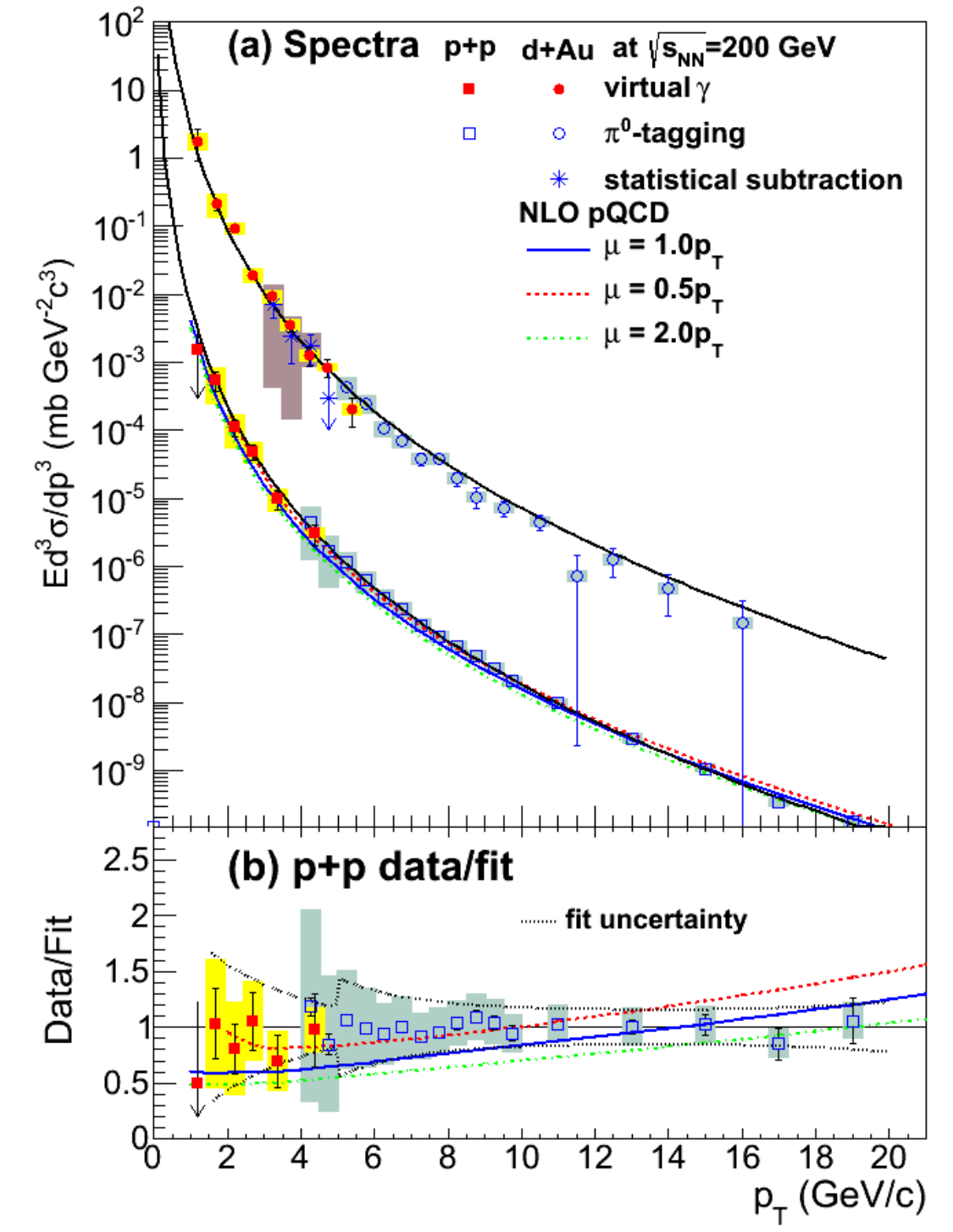}
           \includegraphics*[width=65mm, height=70mm] {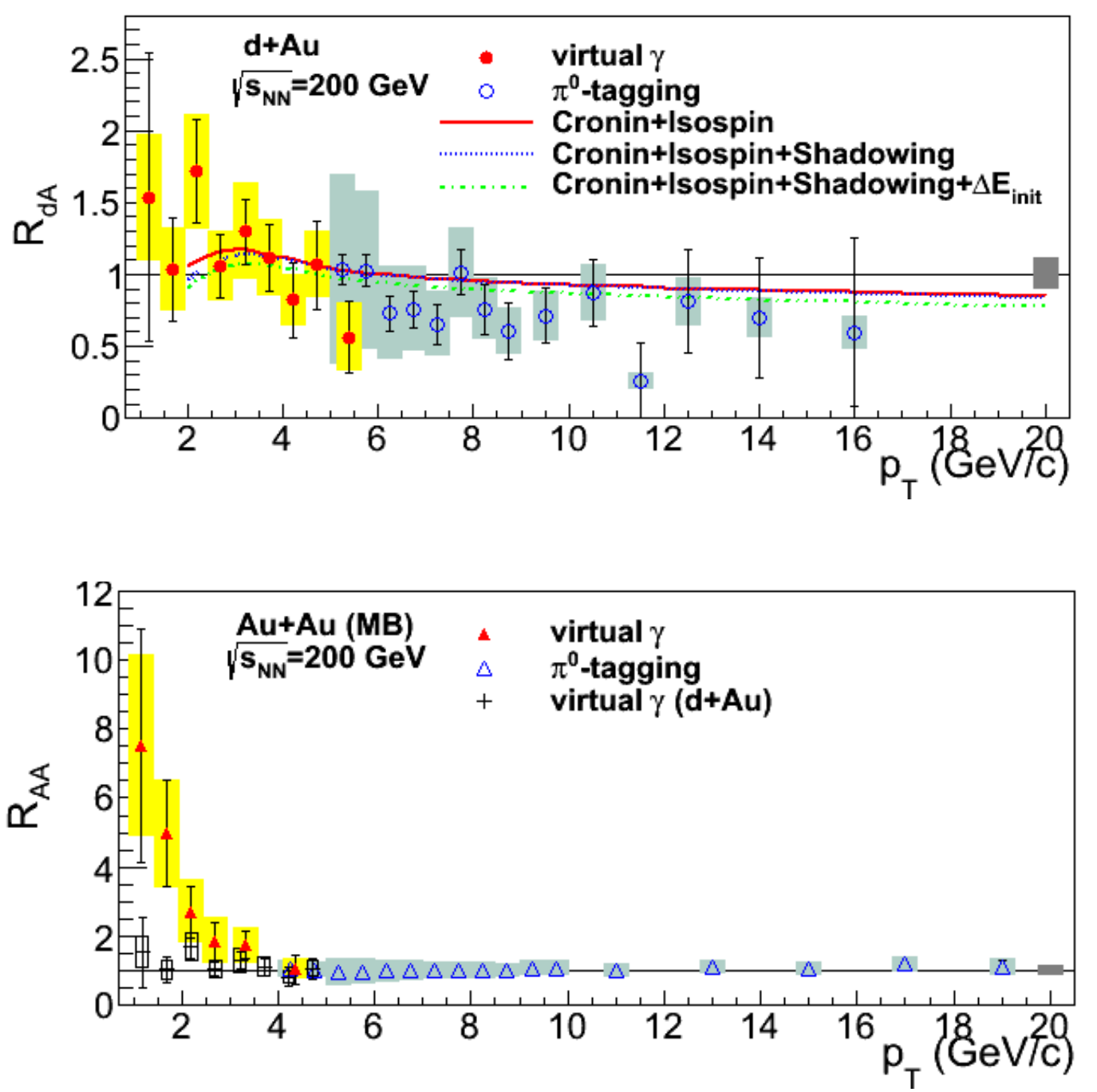}             
           \caption{Left panel: Direct photon invariant spectra in p+p and minimum bias \dau collisions at \sqnr. The figure shows the fit of the p+p data with an empirical parametrization and the results of a NLO pQCD calculation (see \cite{ppg140-photons-dau} for details). Right upper panel: nuclear modification factor R$_{dA}$ of direct photons in \dau collisions. Right lower panel: Comparison of the nuclear modification factors of direct photons in minimum bias \auau and \dau collisions. \cite{ppg140-photons-dau}.}
           \label{fig:photons-dau}   
\vspace{-0.5cm}
    \end{center}
\end{figure*}

This is a seminal result. In order to further strengthen the interpretation in terms of thermal radiation, PHENIX measured the direct photons in \dau collisions to check whether the observed excess in \auau collisions could originate from cold nuclear matter effects in the initial state. The direct photons in \dau were measured via three independent methods: virtual photons, $\pi^0$ tagging and the statistical subtraction \cite{ppg140-photons-dau}.  All three methods yielded consistent results as shown in the left panel of Fig.~\ref{fig:photons-dau}. The figure also demonstrates that the minimum bias (MB) \dau data are well reproduced by the fit to the p+p data scaled by N$_{coll}$ and that there is no excess of direct photons in the \pt range of 1-3 GeV/c. The difference with the \auau system is better illustrated in the right panels of Fig.~\ref{fig:photons-dau}.  The right upper panel demonstrates that the nuclear modification factor in \dau collisions is consistent with unity whereas the right lower panel  shows a comparison of the \raa in \auau and \dau collisions. The large excess of direct photons observed in \auau at low \pt is not observed in \dau  and one can then conclude that it is not due to initial state effects. This result reinforces the interpretation of the \auau excess as thermal radiation from the sQGP.

\section{Direct photons v$_2$}
\begin{figure*}[h!]
    \begin{center}
           \includegraphics*[height=65mm] {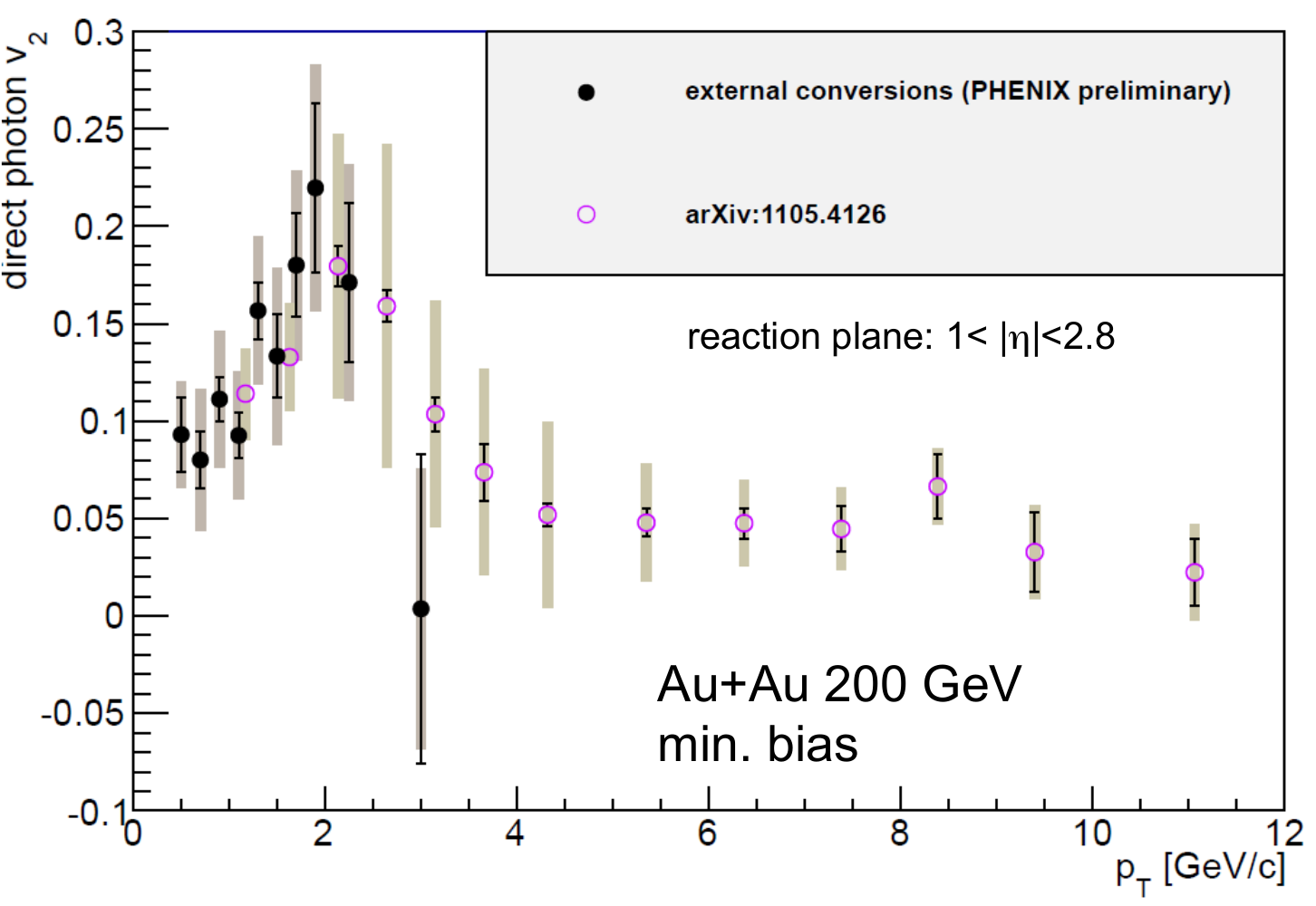}  
          \vspace{-0.4cm}           
           \caption{Direct photon v$_2$ published result (open circles) \cite{ppg126-v2-photons} in minimum bias \auau collisions at \sqnr, compared to the new result (solid points) obtained in an independent analysis based on external conversions. The vertical error bars represent the statistical uncertainties and the shaded boxes indicate the systematic uncertainties.}
           \label{fig:v2-photons} 
         \vspace{-0.4cm}  
    \end{center}
\end{figure*}

PHENIX has measured the elliptic flow v$_2$ of direct photons in \auau collisions at \sqn= 200 GeV \cite{ppg126-v2-photons}. The magnitude and shape of v$_2$ vs. \pt are very similar to those measured for pions.  A large v$_2$ is observed at \pt $<$ 4 GeV/c where the yield is dominated by thermal photons, whereas at higher p$_T$, where the yield is dominated by prompt photons, v$_2$ is consistent with 0 as illustrated  in Fig.~\ref{fig:v2-photons} (open circles). This is a surprising result. Thermal radiation implies early emission whereas a large v$_2$ implies late emission. No wonder therefore that models have difficulties in simultaneously reproducing the yield and the large  v$_2$ values of direct photons.

The importance of this result prompted PHENIX to seek confirmation by performing a completely independent analysis with different systematic uncertainties based on external conversion photons. The new analysis allowed to extend the \pt range of the measurements down to 0.5 GeV/c and yielded consistent results with the previous one in the \pt range of overlap as illustrated in Fig.~\ref{fig:v2-photons}. This provides an important confirmation of the unexpected large v$_2$ values of low \pt direct photons in \auau collisions.

\section {First dilepton results with the HBD}
PHENIX has performed the first measurements of the \ee pair continuum in p+p \cite{ppg085-dileptons-pp} and \auau collisions \cite{ppg088} at \sqnr. The p+p results are well reproduced by the cocktail of known sources whereas the \auau data show a strong enhancement at low masses m =0.15-0.75 GeV/c$^2$ with respect to a similar cocktail adjusted to \auau collisions. The enhancement is concentrated at the most central  0-20\% collisions and reaches a factor of $7.6 \pm 0.5^{stat} \pm 1.3 ^{syst} \pm 1.5 ^{model}$ for the 0-10\% centrality bin and  $4.7 \pm 0.4^{stat} \pm 1.5 ^{syst} \pm 0.9 ^{model}$ for MB collisions.  

In spite of numerous attempts, this result remains  a challenge for theory. In particular, all models that successfully reproduced the low mass pair enhancement observed  by the CERES and NA60 experiments at the SPS energies fail to reproduce the PHENIX result ֿֿ\cite{gale-rapp-tserruya} raising the speculation of a new, yet unidentified, source of dileptons. 

The PHENIX result is limited by large uncertainties, both statistical and systematic, due to the large combinatorial background and the resulting low signal to background ratio S/B and low cocktail to background ratio C/B. The latter represents an objective figure of merit for the sensitivity of the measurement to the known hadronic sources in the absence of any enhancement. For the MB data, the S/B ratio is $\sim$1/200 and the C/B is $\sim$1/1000. In order to qualitatively improve the measurement of low-mass dileptons, PHENIX embarked on an upgrade program. A Hadron Blind Detector (HBD) was developed with the purpose to reduce the combinatorial background. 

\subsection {The Hadron Blind Detector}
The HBD is a windowless Cherenkov detector operated with pure CF$_4$, in a proximity focus configuration. The detector consists of a 50~cm long radiator directly coupled to a triple GEM detector which has a CsI photocathode evaporated on the top face of the upper-most GEM foil and pad readout at the bottom of the GEM stack. The detector is located around the beam pipe in the field free region generated by operating the inner and outer coils of the PHENIX central arms in the $+-$ configuration. In this mode, the inner coil counteracts the main field of the outer coil creating an almost field-free region close to the vertex, extending out to $\sim$50-60 cm in the radial direction \cite{hbd1, hbd2, hbd3}.

The hadron blindness property of the HBD is achieved by operating the detector in the so-called reverse bias mode. 
In this mode, the entrance mesh is set at a lower negative voltage with respect to the top GEM and consequently 
the ionization electrons produced by a charged particle in the drift region between the mesh and the GEM are mostly repelled towards the mesh \cite{hbd2}. 

The choice of CF$_4$ as both the radiator and the detector gas in a windowless configuration results in an unprecedented large bandwidth extending from $\sim$110 nm given by the CF$_4$ cut-off,  up to $\sim$210 nm given by the threshold of CsI . This large bandwidth translates into a large figure of merit N$_0$ and a large number of photo-electrons per electron . 

The main task of the HBD is to recognize and reject  tracks originating from $\gamma$ conversions or $\pi^0$ Dalitz decays, where only the $e^-$ or the $e^+$ is detected in the central arms.  The strategy is to exploit the fact that the opening angle of electron pairs from these sources is very small compared to the pairs from light vector mesons. In the field-free region, this angle is preserved and the pairs produce two close or overlapping hits. Therefore by applying an opening angle cut or a double amplitude cut one can reject a large fraction of the conversions and $\pi^0$ Dalitz decays, while keeping most of the signal.

The detector was successfully operated in 2009 and 2010 and collected data in p+p and \auau collisions, respectively. The data analysis led to a measured value of N$_0 \sim$330 cm$^{-1}$ by far higher than in any other existing gas Cherenkov counter. 
This translates into 20 photo-electrons per single electron incident in the detector, allowing a good separation between single and double hits, a key requirement of the HBD to identify pairs from conversions and $\pi^0$ Dalitz decays \cite{hbd3}. 
The preliminary results shown below exhibit  an improvement of the S/B of $\sim$5 with respect to the previous measurements without the HBD. 

\subsection {Analysis details}
For the p+p data analysis a procedure very similar to the one described in \cite{ppg088, ppg085-dileptons-pp} was used. Significant changes were made for the \auau data analysis. In particular, two parallel and independent analysis streams were followed. The two streams yielded results that are in agreement within their uncertainties, thus providing a crucial consistency check.  A full account of the two streams is far beyond the scope of these proceedings. Only a few aspects and details of the analysis chain that yielded the best results in terms of signal efficiency are described here.

In this first pass through the data we chose to apply strong run QA and strong fiducial cuts. This has the benefit of homogenizing the response of the central arm detectors over time thereby reducing the number of corrections necessary to account for time variations of the various subsystems. On the other hand this approach involves a large price in statistics and pair reconstruction efficiency. 

CF$_4$ has a strong scintillation line at 160 nm, right in the middle of the HBD sensitive bandwidth. This results in a large occupancy of the detector for central collisions that requires an underlying event subtraction before attempting  to identify the electrons hits in the HBD. Two different methods were used for this subtraction yielding consistent results. The HBD single electron detection efficiency achieved was  $>$90\% for the 40-60\% centrality bin gradually decreasing to 65\% for the most central bin of 0-10\% while rejecting 90\% of the fake tracks in both cases.

The HBD provides strong electron identification capability  in addition to the RICH and EMCal detectors of the central arms. Instead of applying a series of independent one-dimensional cuts that ultimately result in a poor efficiency, we used a neural network for electron identification \cite{neural-network}. The neural network was trained on samples of HIJING events covering all centralities. A siginificant improvement of the global single electron track efficiency (of $\sim$ 50\%  for the peripheral 60-92\% events and $\sim$ 20\%  for the most central 0-10\% events) was obtained with the neural network as compared to the standard electron identification cuts. A neural network was also used to inspect the vicinity of electron hits and decide whether the HBD signal corresponds to a single or a double electron hit.

The combinatorial background was subtracted in two separate steps. First the background resulting from the random combination of $e^-$ and $e^+$ in the same event was subtracted using a mixed event technique. The normalization of the unlike mixed event backround is obtained by normalizing the like-sign mixed event spectrum to the like-sign foreground spectrum in a phase space region which is free from correlations (mass $>$ 0.55 GeV/c$^2$ to avoid cross pairs \cite{ppg088}  and opening angle $<$ 120$^o$ to avoid jet correlations).
After this first step,  the foreground unlike-sign spectrum contains both correlated background pairs and signal pairs and the foreground like-sign spectra contain only correlated background pairs. In the second step, the residual like-sign spectra, corrected for the acceptance difference between $++, --$ and $+-$ pairs, are subtracted from the residual unlike-sign spectrum to obtain the raw signal yield \cite{ppg088}. 
This approach is relatively simple but results in a factor of $\sqrt{2}$ larger statistical error as opposed to the case in which the various background sources are determined a priori and subtracted one after the other, without relying on the measured like-sign spectra.

The raw signal is then corrected for acceptance and reconstruction efficiency to yield the invariant dilepton mass spectrum within the ideal acceptance of the PHENIX detector. Conservative systematic errors were assigned to the various steps of the analysis and in particular to the combinatorial background subtraction.

The \auau data were analyzed in the same five centrality bins as in the previoulsy published PHENIX dilepton results \cite{ppg088}. In these proceedings we are presenting the results for peripheral  (60-92\%), semi-peripheral (40-60\%) and semi-central (20-40\%) events. Results from the two most central bins (0-10\% and 10-20\%) are not yet available. 

\subsection{Cocktail of hadronic sources}
The results are compared to a cocktail containing all the known hadronic sources. The EXODUS event generator \cite{ppg088} is used to simulate the photonic sources (Dalitz decays of light neutral  mesons: $\pi^{0}$, $\eta$, $\eta$$'$ $\rightarrow e^+e^-\gamma$ and $\omega \rightarrow e^+e^-\pi^0$) and the non-photonic sources (di-electron decays of the light vector mesons: $\rho$, $\omega$, $\phi \rightarrow e^+e^-$).  
The correlated pairs from semi-leptonic decays of heavy flavor (charm and bottom) mesons are generated using the MC@NLO package \cite{mc-at-nlo} which calculates the initial hard scattering at the next to leading order. 
After generating the various sources, the cocktail is filtered through the ideal acceptance of the PHENIX detector and smeared with the detector resolution.  

$\pi^{0}$ is the dominant electron source and also the fundamental input for EXODUS. For each centrality bin, we use $\pi^0$ \pt spectra based on PHENIX measurements of neutral and charged pions. For other mesons, the shape of the  \pt distribution is obtained assuming m$_T$ scaling. The absolute normalization is provided by  the meson to  $\pi^0$ ratio at high \pt \cite{ppg088}. 

The  J/$\psi$ line shape for the p+p cocktail is obtained from a full Monte Carlo simulation where J/$\psi$ decays into \ee are propagated through the full chain of reconstruction and analysis cuts using a GEANT based simulation of the PHENIX detector.  
For the J/$\psi$ in \auau, the shape and yield are taken from the p+p measurement, the latter being scaled by N$_{coll}$ and the measured nuclear modification factor \raa  \cite{ppg068-jpsi-raa}. Therefore the comparison cocktail vs data for the J/$\psi$ in \auau is reduced to a consistency check between the previously published and the present J/$\psi$ yields, without and with HBD, respectively.

\subsection {Results}
\begin{wrapfigure}{1}{8cm}
           \includegraphics*[width=80mm]{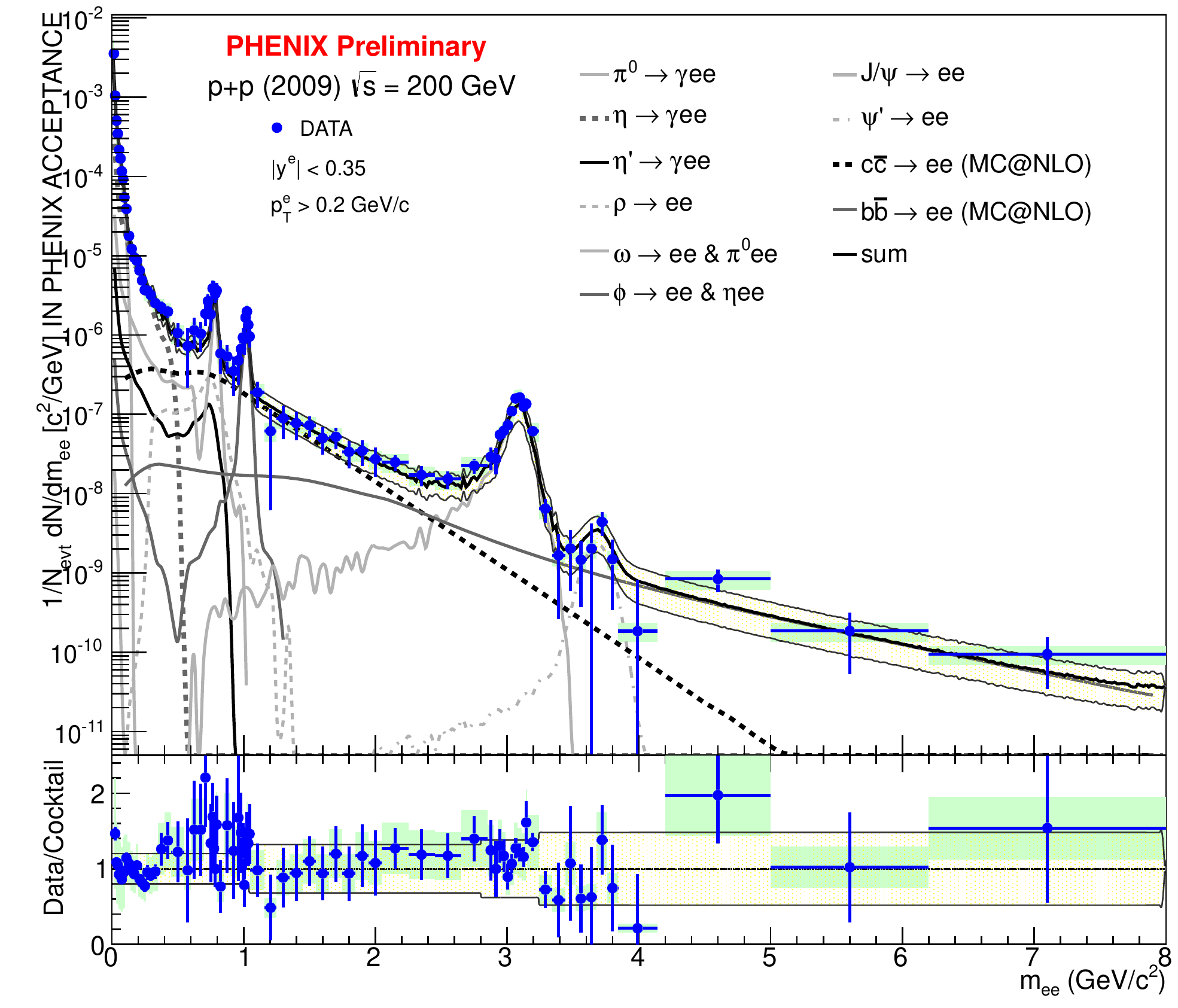}  
           \vspace{-0.6cm}           
           \caption{Invariant mass spectrum of \ee pairs in p+p collisions at \sq = 200 GeV measured in the 2009 run with the HBD. The data points show statistical (vertical bars) and systematic (shades) uncertainties separately. The data are compared to the expected yield from a cocktail of the known hadronic sources. The shaded band represents the systematic uncertainties of the cocktail. The bottom panel shows the data to cocktail ratio.}
           \label{fig:mass-spectrum-pp}
\end{wrapfigure}

Figure \ref{fig:mass-spectrum-pp} shows the invariant mass spectrum of \ee pairs within the PHENIX detector acceptance measured in the 2009 p+p run with the HBD. The results are compared to the expected yield from the cocktail of known sources. The bottom panel shows the ratio of data to cocktail and demonstrates a good  understanding of the measured spectrum. The present result is fully consistent with the previously published one without the HBD \cite{ppg085-dileptons-pp}
and provides a crucial proof of principle for 
usage of the HBD. This analysis is now being repeated using all the tools of the new analysis chain developed for the \auau data analysis in order to fully assess the benefit of these tools.
 
\begin{figure}
    \begin{center}
           \includegraphics*[width=43mm] {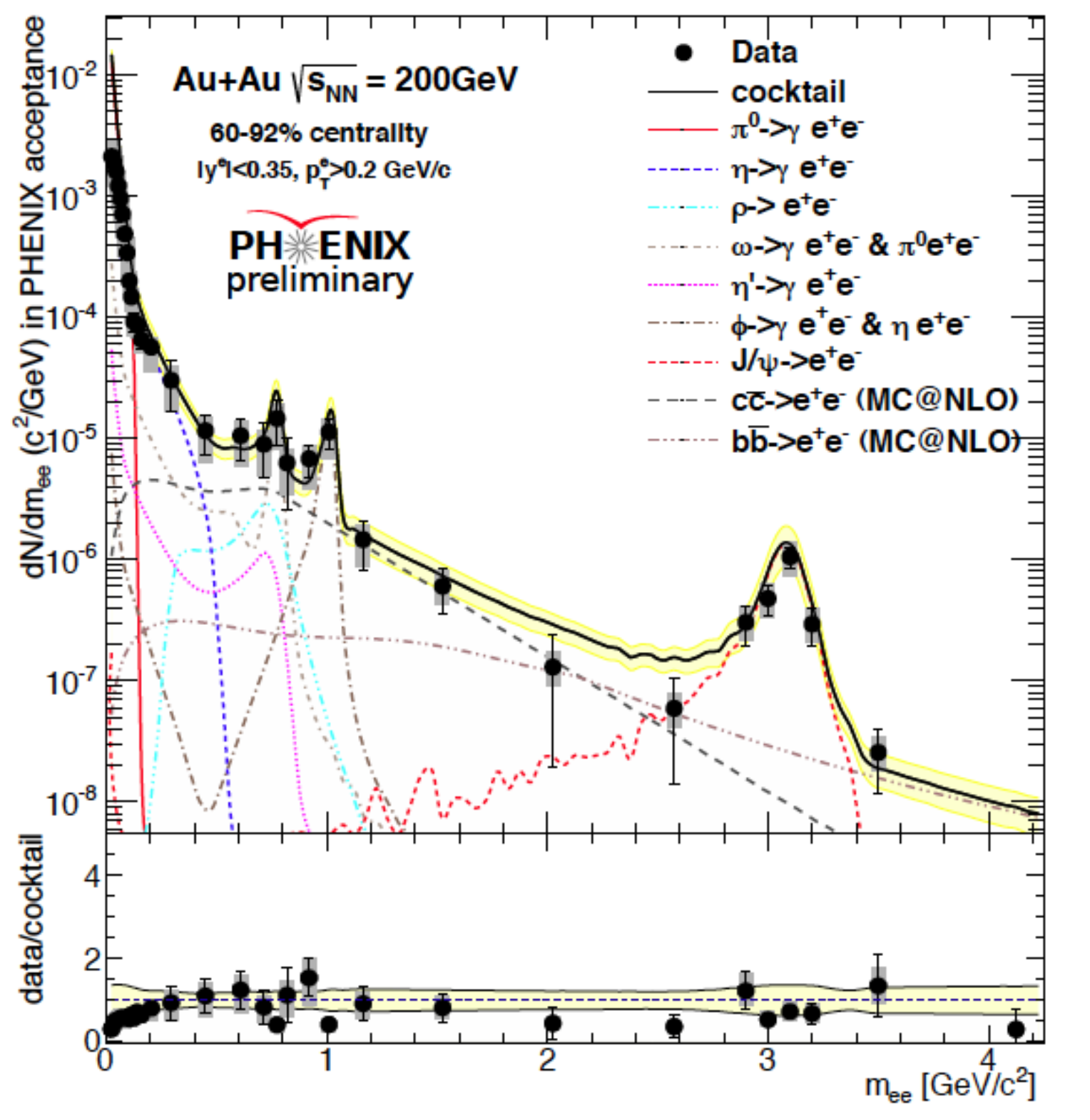}
           \includegraphics*[width=43mm] {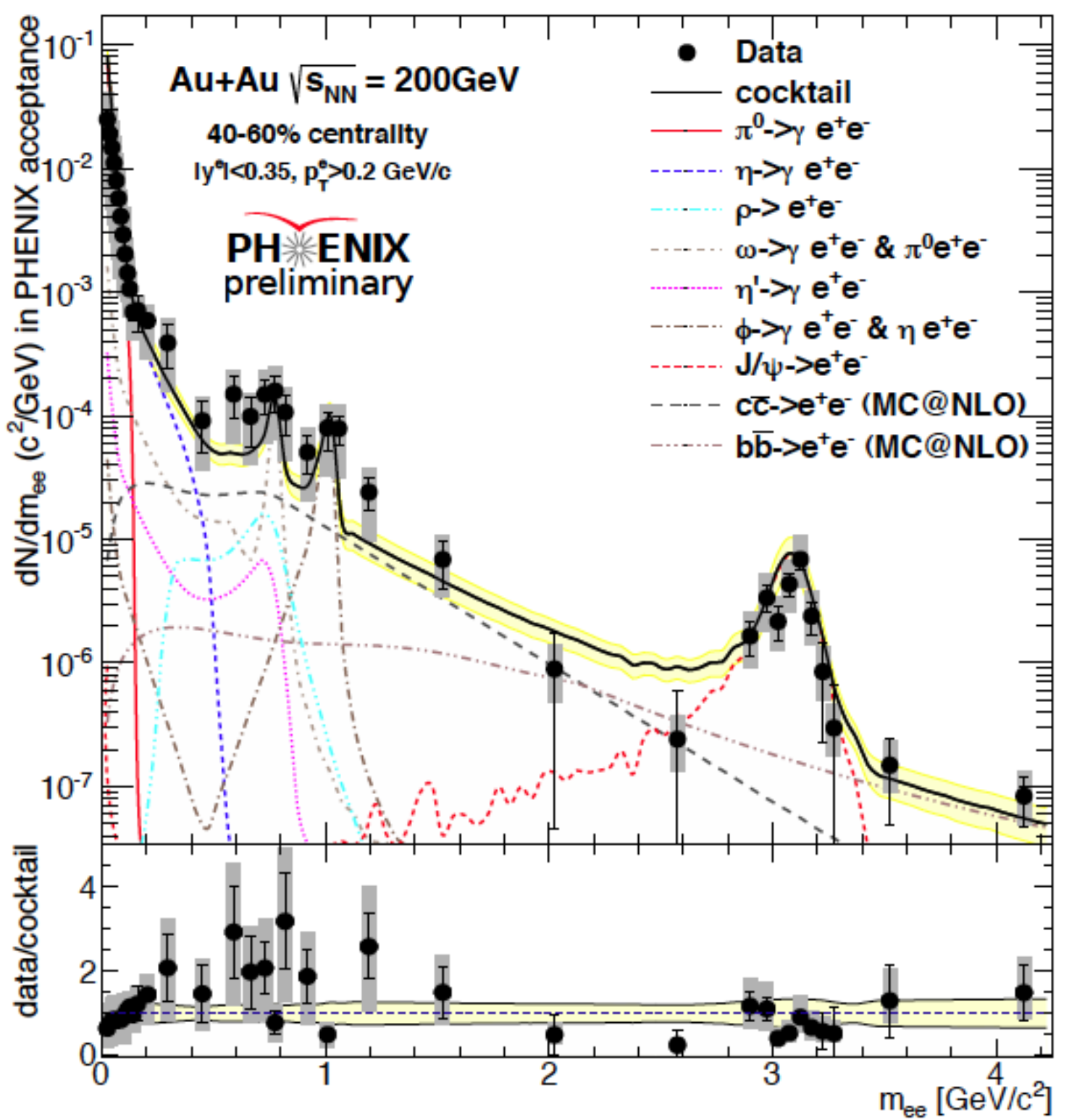}
           \includegraphics*[width=43mm] {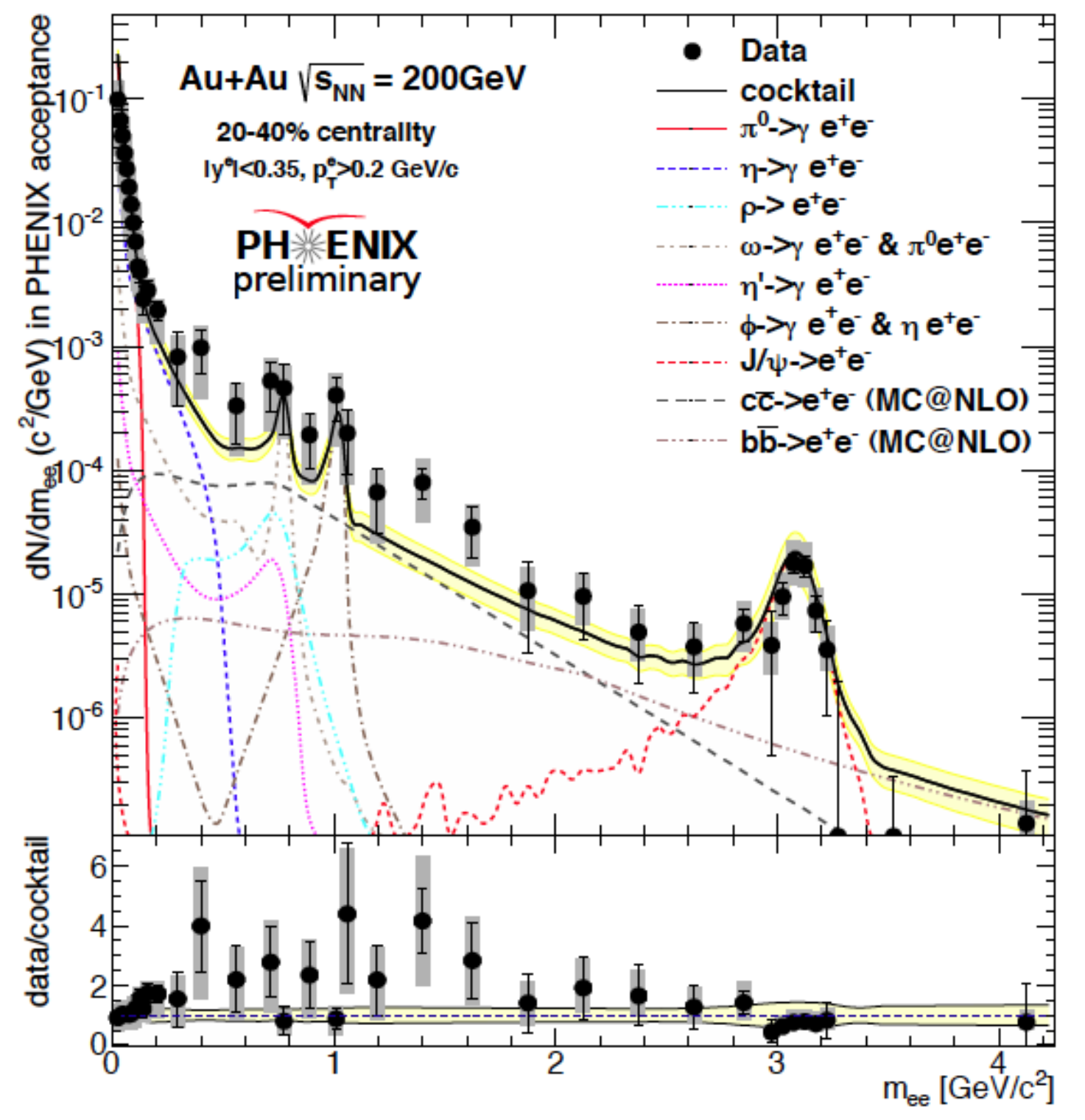}
           \caption{Invariant mass spectrum of \ee pairs in \auau collisions at \sqnr measured in the 2010 run with the HBD in three centrality bins.  The data points show statistical (vertical bars) and systematic (shades) uncertainties separately. 
The shaded band represents the systematic uncertainty of the cocktail.}
           \label{fig:mass-spectra-auau}    
\vspace{-0.7cm}
    \end{center}
\end{figure}

Figure \ref{fig:mass-spectra-auau} shows the invariant mass spectra of \ee pairs within the PHENIX detector acceptance measured in \auau collisions in the 2010 run with the HBD, in three centrality bins: 60-92\% (left panel), 40-60\% (middle panel) and 20-40\% (right panel). The results are compared to the expected yield from the cocktail of known sources, briefly described in the previous section. The bottom panels show the ratio of data to cocktail. In the most peripheral bin the measured spectrum is in quite good agreement with the expected yield. For semi-peripheral and semi-central events there seems to be a small enhancement of the measured yield with respect to the cocktail that increases with centrality, both in the low-mass region (LMR, m = 0.15 - 0.75 GeV/c$^2$) and the intermediate mass region (IMR, m = 1.2 - 2.8 GeV/c$^2$) . However, no clear statement can be made at this stage with the present level of very conservative errors.

\begin{figure*}[h!]
    \begin{center}
           \includegraphics*[width=65mm] {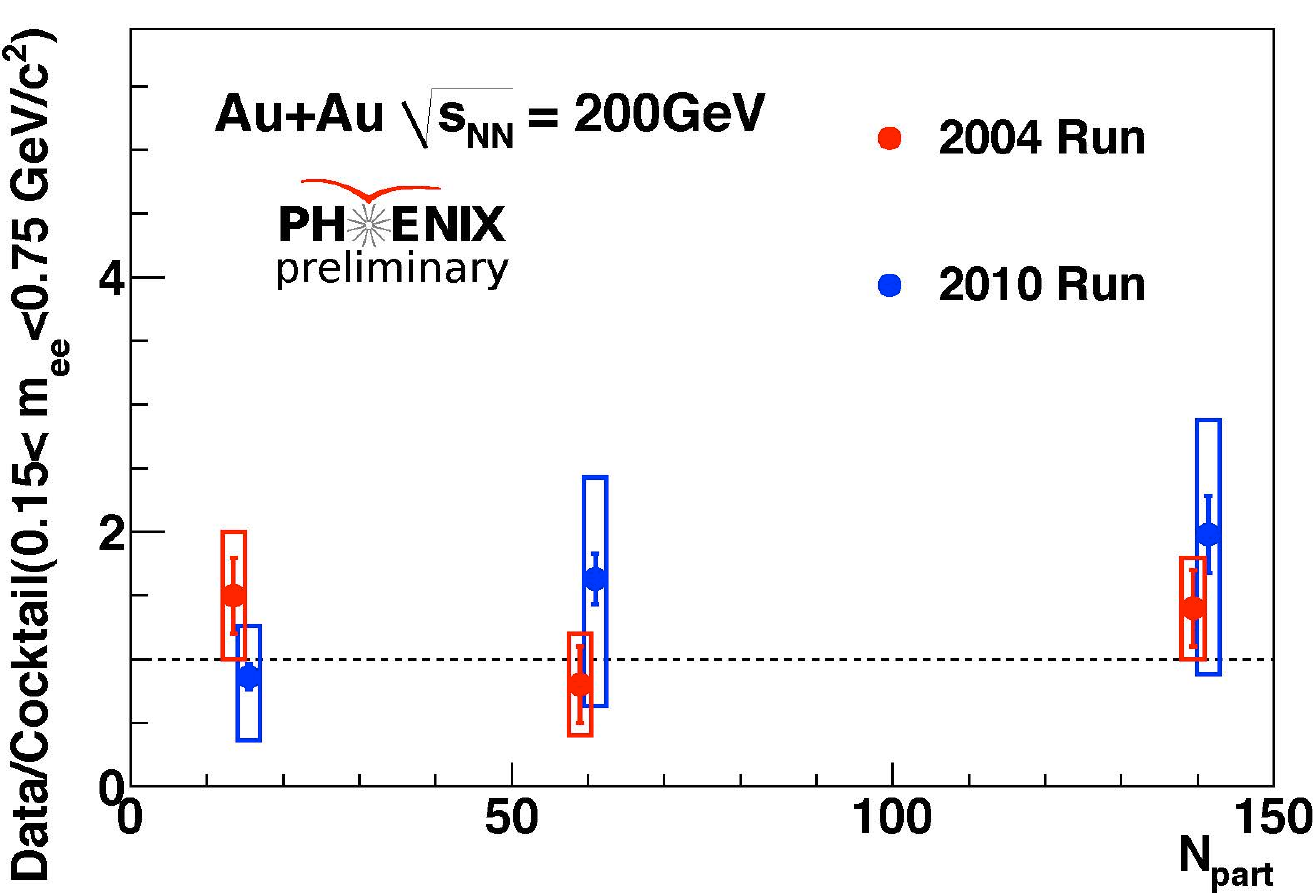}
           \includegraphics*[width=65mm] {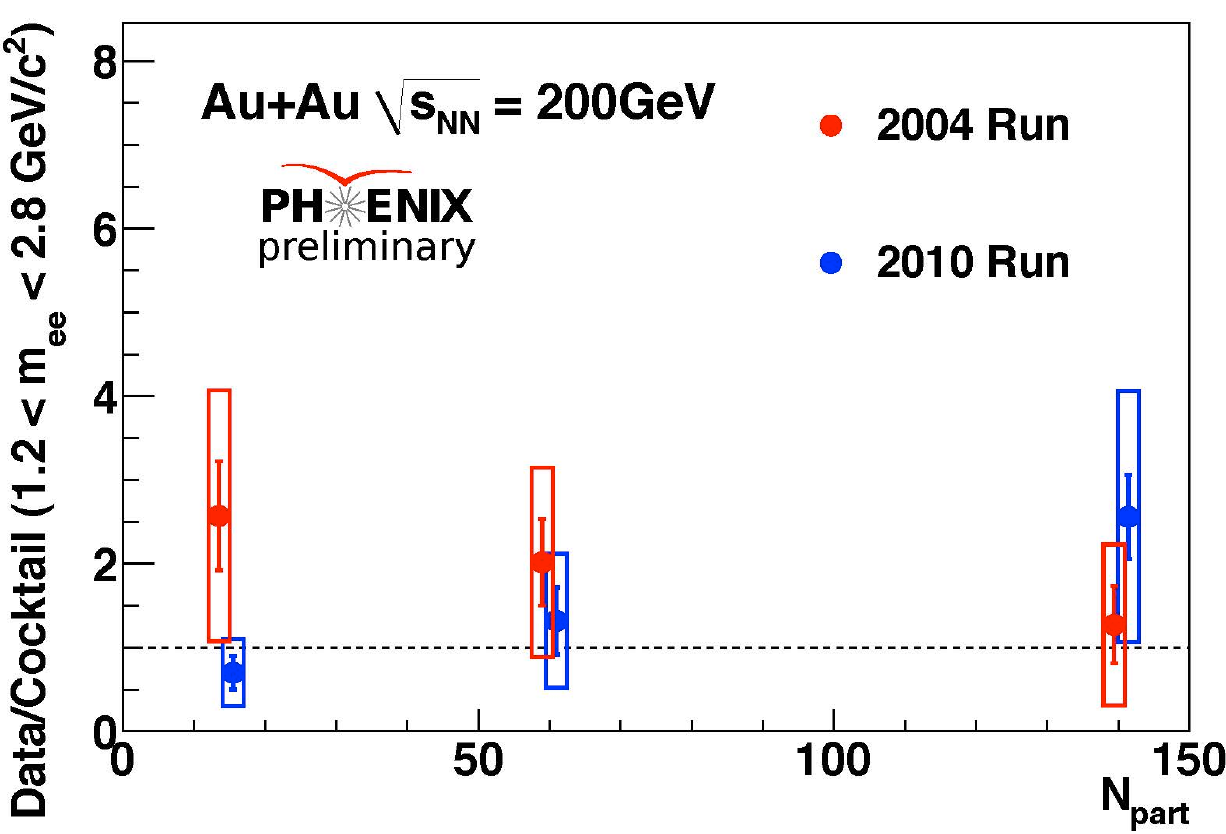}             
           \caption{Comparison of the data/cocktail ratio with (2010 run) and without (2004 run) the HBD of the integrated yield in the LMR (left) and the IMR (right) as function of N$_{part}$.}
           \label{fig:run4-run10} 
\vspace{-1.0cm}  
    \end{center}
\end{figure*}

There are significant differences between the 2010 run with the HBD and the 2004 run without it, that prevent a direct comparison of the results. In the 2010 run the central coils were operated in the $+-$ configuration, as opposed to $++$ in 2004, resulting in larger acceptance of low \pt tracks. There is more radiative tail in the J/$\psi$, $\phi$ and $\omega$  line shapes due to the HBD material in front of the central arms. There is also a difference in the cocktails. In the 2010 run, the open heavy flavor contribution is calculated using the MC@NLO package whereas in the 2004 data analysis we used PYTHIA. This results in a calculated yield in the IMR  which is $\sim$16\% larger with the MC@NLO than with PYTHIA. To account for these differences, we compare in Fig.~\ref{fig:run4-run10}  the data/cocktail ratio for the integrated yield in the LMR
and the IMR as obtained in the 2004 and 2010 runs for the three centrality bins presented here. The 2004 PYTHIA yield in the IMR is scaled up by 1.16 to account for the difference with the MC@NLO calculations. The figure shows consistent results between the two runs within their uncertainties.

\section{Summary and outlook}
PHENIX is pursuing its systematic study of electromagnetic probes exploiting the flexibility of the RHIC facility. Recent achievements presented here, include final results on \raa of direct photons \cite{ppg139-raa-photons}, independent confirmation of the large v$_2$ values of direct photons and first measurement of direct photons in \dau collisions. The preliminary dilepton analysis using the HBD,  with strong QA cuts and conservative error estimates, yielded  consistent results with those previously published \cite{ppg088, ppg085-dileptons-pp}. The final analysis is expected to show a significant improvement in the statistical and systematic uncertainties and to include the two most central bins.

\end{document}